**All-optical autoencoder machine learning framework using linear diffractive processors**

*Peijie Feng, Yong Tan, Mingzhe Chong, Lintao Li, Zongkun Zhang, Fubei Liu, Yongzheng Wen\* and Yunhua Tan\**

P. Feng, M. Chong, Z. Zhang, Y. Tan

School of Electronics, Peking University, Beijing, 100871, China

E-mail: tanggeric@pku.edu.cn

Y. Tan, L. Li, F. Liu, Y. Wen

State Key Laboratory of New Ceramics and Fine Processing, School of Materials Science and Engineering, Tsinghua University, Beijing 100084, China & Research Center for Metamaterials, Wuzhen Laboratory, Jiaxing 314500, China

E-mail: wenyzheng@tsinghua.edu.cn

**Abstract**

Diffractive deep neural network (D$^2$NN), known for its high speed and strong parallelism, has been widely applied across various fields, including pattern recognition, image processing, and image transmission. However, existing network architectures primarily focus on data representation within the original domain, with limited exploration of the latent space, thereby restricting the information mining capabilities and multifunctional integration of D$^2$NNs. Here, we propose an all-optical autoencoder (OAE) framework that linearly encodes the input wavefield into a prior shape distribution in the diffractive latent space (DLS) and decodes the encoded pattern back to the original wavefield. By leveraging the bidirectional multiplexing property of D$^2$NN, the OAE models function as encoders in one direction of wave propagation and as decoders in the opposite direction. We further apply the models to three key areas: image denoising, noise-resistant reconfigurable image classification, and image generation. Proof-of-concept experiments have been conducted to validate numerical simulations. Our OAE framework fully exploits the potential of latent representations, enabling a single set of diffractive processors to simultaneously achieve image reconstruction, representation, and generation. This work not only offers fresh insights into the design of optical generative models but also paves the way for developing multifunctional, highly integrated, and general optical intelligent systems.

## 1. Introduction

Deep learning, a subset of machine learning, has achieved remarkable performance in various fields over the past decade, including digital image processing[1], natural language processing[2], pattern recognition[3], intelligent device design[4], and the discovery of physical laws[5]. By employing multilayered artificial neural networks with plenty of parameters to simulate complex brain processes, deep learning enables computers to automatically extract features and learn from extensive datasets. Among the techniques of deep learning, generative models have garnered significant attention recently due to the success of large language models (LLMs).[6] Typical models such as autoencoders (AEs)[7], generative adversarial networks (GANs)[8], and transformers[9], essentially adhere to the encoding-decoding framework. By designing appropriate model structures and loss functions, a set of encoders and decoders is trained to extract enhanced data representations in latent space from raw data, revealing its inherent characteristics. By further leveraging the latent representations, generative models have demonstrated significant application potential in data generation, such as artistic content creation and speech synthesis.[10,11]

However, with the explosive growth in model scale and data volume, the bottleneck of computational power and storage capacity imposed by Moore's Law, coupled with the vast power consumption brought by network operations, hinder further advancements in deep learning model performance.[12] To address these limitations, optical neural networks (ONNs), leveraging the unique properties of light to achieve high speed, ultra-wide bandwidth, and energy-efficient computations, are proposed as a promising alternative.[13-15] The deep diffractive neural network ($D^2NN$) is a prominent example of ONNs, which utilizes light diffraction to establish connections and diffractive layers to modulate information.[16] Moreover, metasurfaces provide an ultra-thin optical platform to implement diffractive layers. [17,18] Composed of numerous diffractive neurons, they enable subwavelength resolution control of light phase and amplitude, supporting the development of miniaturized and integrated intelligent optical systems.

Recent advances in $D^2NN$ have demonstrated its potential for efficient execution of deep learning tasks in the photonic domain. As for pattern recognition, $D^2NN$s have been applied to image classifications, [19-21] sorting orbital angular momentum (OAM) beams, [22] and estimating the directions of arrival (DOA) of waves. [23,24] These applications require $D^2NN$s to accurately recognize different types of input wavefields and focus the output on the corresponding locations to achieve optical recognition. In the field of light computation, D$^2$NNs are utilized to perform linear matrix calculations, [25] digital logic operations, [26] and trigonometric functions, [27] where they are trained to fit specialized calculators. Diffractive image processors implemented by D$^2$NNs, such as image denoisers, [28] unidirectional imagers, [29,30] and super-resolution imagers[31] have been widely studied to tackle the high loss and complexity of traditional electronic systems. Beyond the above, there are also researches exploring the multiplexing mechanisms[32-35] and superior network structures[20,21,36] to enhance the parallelism, reconfigurability, and scale of the models.

However, these works focus solely on the end-to-end representation of data in the original domain. Specifically, the input optical field carrying the original information passes through the D$^2$NN and directly forms an artificially designed target wavefield at the output plane to perform a specific task. The latent data space, where data can be represented more concisely, is severely neglected. Although several studies have proposed optoelectronic encoder-decoder models that can transform the original information into latent space, [37-39] the cascading of electronic and optical modules imposes significant burdens due to optical-to-electrical conversion and mismatches. Recently, an all-optical VAE model implemented by a spatial light modulator (SLM) has been reported to optimize the traditional fiber optic communication system[40]. The well-trained optical encoder and decoder are utilized to replace the optoelectronic hybrid transmitter and receiver, respectively. However, limited by the scenario, only the compression capability and reconstruction operation of the latent space are realized. All in all, current D$^2$NNs have not fully explored the potential for representation and application of the latent data space, which makes models tailored for specific tasks, lacking integration and versatility.

In this article, we propose an all-optical autoencoder (OAE) machine learning framework, which consists of a linear encoder and a linear decoder that share a set of diffractive layers, achieving weight sharing. Benefiting from the bidirectional multiplexing property of the D$^2$NN (see Note S1, Supporting Information), Our framework acts as an encoder that transforms original image to encoded patterns in the diffractive latent space (DLS) along $FOV_\text{I} \to FOV_\text{II}$ direction, while as a decoder that reconstructs the input images from DLS along $FOV_\text{II} \to FOV_\text{I}$ direction. The encoded patterns are regularized by different prior shape distributions, giving rise to various modes of the OAE framework. The single optical autoencoder (SOAE) model and the multiple optical autoencoder (MOAE) model are basic modes. The former encodes the input images from distinct classes into the same region, while the latter encodes them into separate and class-specific regions. Extended modes further design other six shapes of the encoded region in the DLS, which are analyzed in Note S2 (Supporting Information). Additionally, we apply the SOAE model and MOAE model in three domains. To achieve image denoising, we transmit noisy images bidirectionally along $FOV_\text{I} \rightarrow FOV_\text{II}$ and $FOV_\text{II} \rightarrow FOV_\text{I}$, utilizing the encoder and the decoder in a cascaded manner. This approach allows each diffractive layer to modulate the incident information twice, enhancing the compactness of the diffractive denoiser. We also construct reconfigurable classifiers that can adapt to different datasets by altering pluggable classification layers behind the fixed encoder. Owing to the concise features extracted by the encoder, the classifiers demonstrate improved noise resistance performance. Finally, image generation, including hologram generation and conditional hologram generation, is achieved using the decoders of the SOAE and MOAE models, respectively. Conditional hologram generation here means that the category of the generated image is controlled by the spatial position of the encoded region, while the details of the image are determined by the encoded pattern. It remarkably improves the channel capacity and flexibility of metasurface holography.

To validate the proposed OAE framework and its applications, we manufactured several sets of diffractive layers using 3D printing technology and conducted proof-of-concept experiments in the terahertz (THz) band. The experiments successfully proved the effectiveness of numerical simulations. Although our experiments were designed in the THz band, this framework can be universally extended to lower millimeter-wave bands or higher optical bands. For one thing, our framework employs the bidirectional multiplexing mechanism, using only a set of weights for self-encoding and decoding, broadening the scope of the term auto. For another, the prior shape distribution of the DLS provides a new optical coding dimension for encoding design. Three application scenarios demonstrate that the framework can simultaneously achieve data reconstruction, representation, and generation, embodying compactness and multifunctional integration. Our work will facilitate the design of optical generation models and versatile optical intelligent systems, which can be further applied to more typical tasks, such as image encryption, obstacle-avoiding optical communication, and off-axis holography.

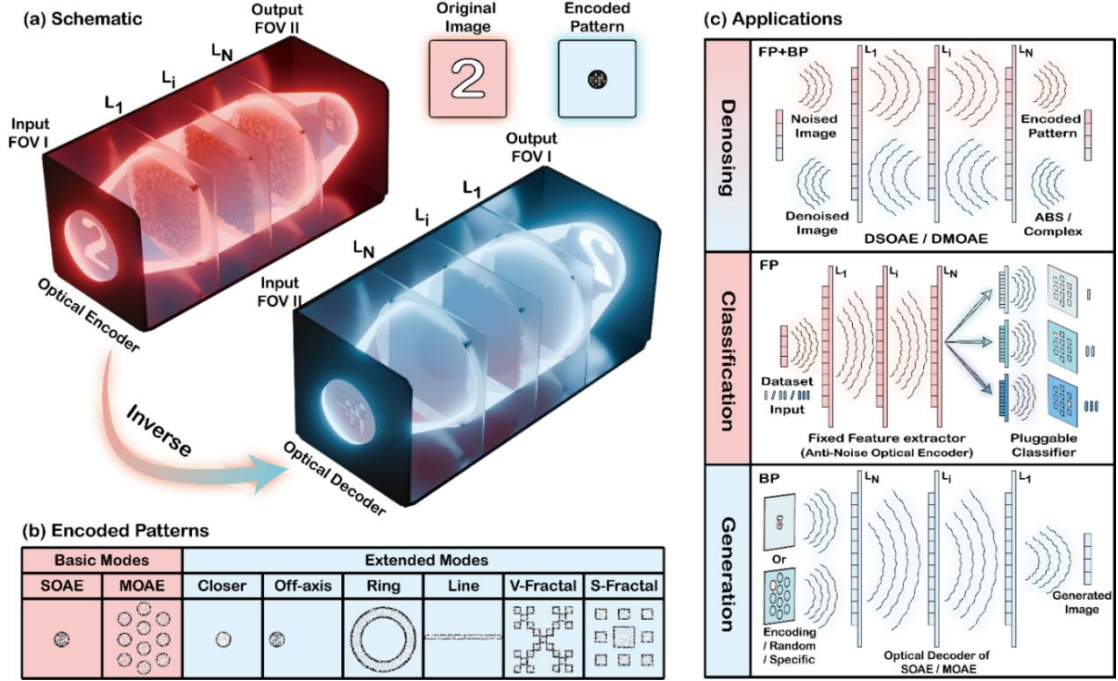

**Figure 1.** System schematic and applications of the all-optical autoencoder machine learning framework. (a) System diagram of the OAE framework. It uses only one set of diffractive layers ($L_1, ..., L_i, ..., L_N$) to separately implement the optical encoder and optical decoder in two directions. (b) DLS encoding pattern design. The basic modes include SOAE and MOAE. The abbreviation S represents the single encoding region, while M denotes multiple encoding regions. The extended modes explore six different encoded patterns in DLS. (i.e., Closer: Altering the distance between $FOV_{II}$ and $L_N$ closer. Off-axis: Moving encoded region bias from the center. Ring: Making encoded patterns distribute in a ring. Line: Arranging encoded patterns along the line in a one-dimensional order. V-Fractal and S-Fractal: Designing the shape of encoded region as Vicsek and Sierpinski carpet fractal[41]) (c) Three applications of the SOAE and MOAE models: image denoising (bidirectional transmission), noise-resistant reconfigurable image classification ($FOV_I \to FOV_{II}$), and hologram generation ($FOV_{II} \to FOV_I$).

## 2. Results

In this article, we consistently define the forward direction as the light path from field of view Ⅰ ($FOV_I$) to $FOV_{II}$, while the backward direction is identified as reverse path ($FOV_{II} \to FOV_I$). $FOV_I$, also called the original space, is shared by the original field and the reconstructed field. The encoding region in $FOV_{II}$, determined by the prior shape distribution, is referred to as the DLS. Our OAE framework functions as an encoder in the forward direction and as a decoder in the backward direction (see **Figure 1a**), achieving self-encoding and decoding through a bidirectional multiplexing mechanism. It is noteworthy that although our DLS and the latent space of electronic networks can both efficiently represent information in an alternative space, the DLS here is restricted to linear operations and exists in the form of a two-dimensional optical field, preventing it from resembling the latent space of electronic models.

Without the loss of generality, the information in original space and DLS discussed in this section are represented using the amplitudes of complex fields. Specifically, the input images and reconstructed images in the original space, along with the encoded patterns in the DLS, are represented by the amplitudes of wavefields in their respective regions. Furthermore, our model can also convey information in the original space using the intensity or phase of light (see Figure S19, Supporting Information) and represent encoded patterns using complex fields (see Note S3, Supporting Information). MNIST, EMNIST, and Fashion datasets were utilized for model training and inference throughout this paper.

## 2.1 Performance of the MOAE model and the SOAE model

Two basic modes of the OAE framework, the SOAE and MOAE models, are first illustrated here. The encoded patterns of these two models are both a circle with a radius of 5, whose dimension is significantly smaller than that of original images ($64 * 64$), resulting in a compression ratio of approximately 52. The SOAE model employs only a single encoding region for different categories of images, while the MOAE model assigns distinct encoding regions to different classes of images, mapping each class to its corresponding region. (see Figure 1b). During model training, the original images first pass through the model in the forward direction. Subsequently, the amplitudes of the wavefield in the DLS are extracted as the encoding pattern. Finally, this encoding pattern is processed in the reverse direction through the model to obtain the reconstructed images. To assist the model in achieving effective encoding and decoding reconstruction while ensuring system efficiency, the loss function $\mathcal{L}_{OAE}$ is designed and defined in Methods part. We also employed three metrics: Structural Similarity Index (SSIM), Peak Signal-to-Noise Ratio (PSNR), and efficiency ($\eta$) to comprehensively evaluate the trained models. Among them, SSIM and PSNR assess the quality of the reconstructed images, whereas $\eta$ evaluates the forward and backward energy transmission efficiency of the models.

Figure 2a shows the phase profiles of the SOAE and MOAE models trained on the MNIST dataset. The SOAE model consists of five equally spaced phase-only modulation diffractive layers (N=5), with phase distributions that exhibit distinct center-focus characteristics, directing the original images toward the central encoding region. In contrast, the MOAE model includes six diffractive layers (N=6), with its phase distributions gradually transitioning from a single center to multiple centers, guiding the input images to converge on ten encoding regions. The encoding and decoding results of these two models are presented in Figure 2b. The SOAE model compresses the image into the encoding region during forward encoding, while enlarges the compressed images to the original size during reverse decoding, which is similar to a convex lens. This lens-like characteristic grants the SOAE model strong generalization capabilities (see Figure S3, Supporting Information). Therefore, our framework utilizes several planar diffraction layers designed by a data-driven deep learning algorithm to create a specialized convex lens, indicating that $D^2NN$ has the potential to extract general physical principles from specific datasets. The MOAE model transforms the original images into ten encoded patterns in the DLS. By reversely passing patterns in proper position through the model, it can also produce well-reconstructed images. The encoded patterns of this model possess encryption characteristics. Figure 2c shows the performance of both models trained on the EMNIST and Fashion datasets. Notably, the SOAE model trained on the Fashion dataset does not exhibit the lens-like characteristics. It is because the high complexity of the dataset makes it difficult for linear $D^2NN$ to extract general physical principles.

Figure 2d shows the convergence curves of the SOAE and MOAE models, indicating that they achieved stable convergence in training losses after several hundred of training epochs. Bar charts of metrics for both models trained on MNIST, EMNIST and Fashion datasets are illustrated in Figure 2e. The SOAE models trained on the MNIST and EMNIST datasets, which exhibit lens-like characteristics, demonstrated high image reconstruction quality and transmission efficiency (SSIM > 0.85, PSNR > 20 dB, and $\eta$ > 50%). In contrast, due to the complexity of the Fashion dataset, the SOAE model trained on it showed the worst image reconstruction quality, with a SSIM of 0.73 and PSNR of 16.27dB. Furthermore, $\eta_F$ of this model was also lower than that of previous two models, at only 10.90%. The performance of MOAE models across the three datasets was relatively similar (SSIM ~ 0.65, PSNR ~ 14.5 dB, $\eta_F$ ~ 4%, and $\eta_B$ ~ 20%). However, when compared to the SOAE models, the overall performance is degraded due to the more complex encoding task of achieving spatial multiplexing for ten encoding regions.

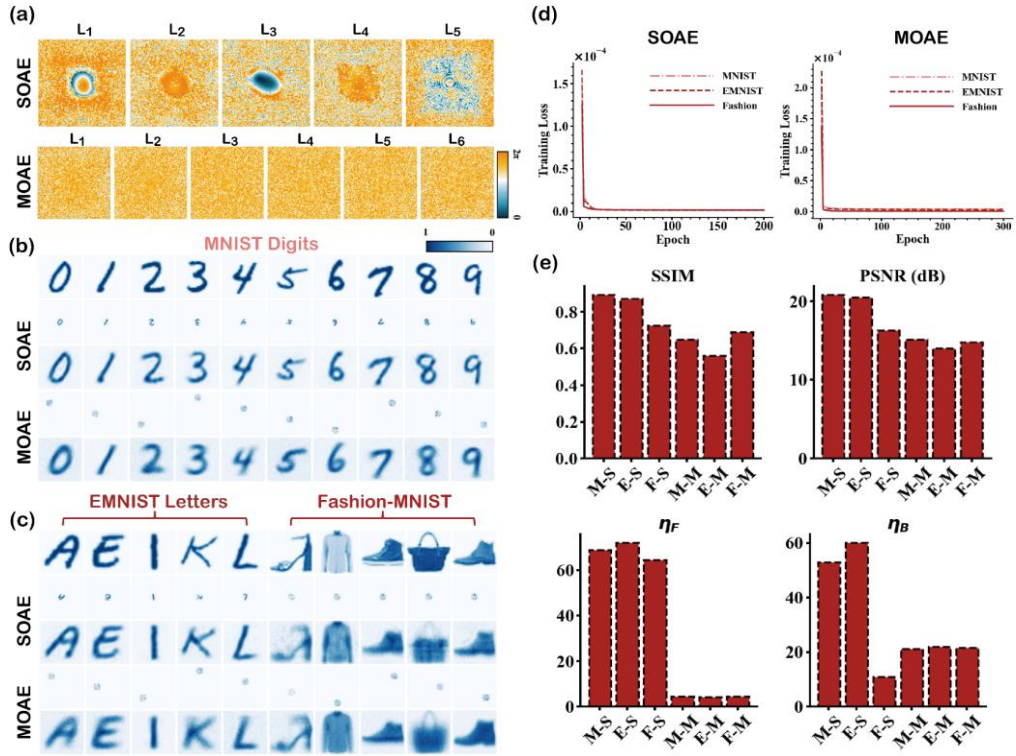

**Figure 2.** The results of the two basic modes, SOAE model and MOAE model. (a) Phase profiles of for SOAE model and MOAE model trained on MNIST dataset. (b) The encoding (2nd row and 4th row) and decoding (3rd row and 5th row) results of two models on the MNIST dataset. The first row shows the ground truth. (c) The encoding and decoding results for five classes of samples from the EMNIST and Fashion datasets using two models. The models were trained on the ten-class datasets. The remaining five classes and model phase distributions are detailed in Figure S2 (Supporting Information). (d) The training convergence curves of the SOAE model and MOAE model. (e) The evaluation metrics for two models (abbreviation S and M) training on the MNIST (M), EMNIST (E), and Fashion (F) datasets.

## 2.2 Application 1: Image denoising

The SOAE model and MOAE model were further employed for three tasks to demonstrate the multifunctional integration characteristics of our framework. In this part, the encoders and decoders of the well-trained SOAE and MOAE models are cascaded to serve as diffractive image denoisers, which could remove the pepper-and-salt noise and uniformly distributed noise in the images. Note S5 (Supporting Information) defines the models for the two types of noise. The noised images first pass through the models forward to obtain the encoded patterns. Subsequently, the patterns propagate backward through the models to generate the denoised images. Additionally, we train the models on a dataset with noise (i.e., noisy input and clean target output) to enhance their denoising capability, resulting in the denoising single optical autoencoder (DSOAE) and denoising multiple optical autoencoder (DMOAE) models.

Figure 3a presents the ability of these two models to eliminate pepper-and-salt noise through typical samples. The parameter $\alpha$ adjusts the noise intensity by controlling the proportion of pixels affected by noise relative to the total number of pixels. For the SOAE model, the PSNRs between reconstructed and denoised images differ by less than 1.6 dB when $\alpha$ is less than 0.5, resulting in high image clarity. The quality of denoised images will rapidly decrease as $\alpha$ increases from 0.5 to 0.7, with PSNR of 15.5 dB at $\alpha = 0.7$. $\eta_F$ also decreases from around 70.3% to around 25.8% with the growth of $\alpha$ because the removed noise carries significant energy, which means that the denoising process primarily occurs during the forward transmission. This indicates that our model can extract useful DLS information from noisy inputs. The denoising behavior of the MOAE model aligns with that of the SOAE model, with both PSNR and $\eta_F$ decreasing as noise intensity increases. The performance of models on removing uniformly distributed noise is demonstrated in Figure 3b. We control the noise intensity by setting its upper bound to $\beta$ times the mean square value of the image. The SOAE and MOAE models can also eliminate this noise following the analysis of pepper-and-salt noise. Moreover, the four additionally trained DMOAE and DSOAE models can effectively enhance the denoising image clarities and the system efficiencies. By comparison, high-intensity noise will add extra background to images processed by the SOAE model and cause images handled by the MOAE model to become blurred. Uniformly distributed noise has a weaker impact on image contours compared to pepper-and-salt noise.

Figure 3c illustrates the PSNR histograms of noisy and denoised MNIST validation sets. The first row presents the PSNR distribution of noisy images, while the second and third rows show the PSNR distribution of denoised images obtained by the labeled models. The median PSNR of the dataset affected by pepper-and-salt noise is 7.81 dB. This value improves significantly to 17.3 dB with the SOAE model and further increases to 18.6 dB using the DSOAE model. The MOAE and DMOAE denoisers respectively enhance the median to 14.4 dB and 14.8 dB. Uniformly distributed noise has a milder interference effect on the images, resulting in a higher median PSNR of 9.94 dB. After processing with the SOAE, DSOAE, MOAE, and DMOAE models, the median PSNR increases to 14 dB, 19 dB, 14.2 dB, and 14.9 dB, respectively. Figure 3d reflects the impact of noise intensity on PSNR and transmission efficiency (average value). By comparing PNSR curves in the four plots and median values of Figure 3c, we can conclude that the SOAE model's ability to remove uniformly distributed noise is weaker than its capacity to eliminate pepper-and-salt noise. However, the DSOAE model effectively compensates for this shortcoming, significantly enhancing the model's denoising capability for the former. The MOAE and DMOAE models perform comparably for both types of noise. Moreover, the maximum PSNR improvement of 10.89 dB is achieved by the DSOAE model at $\alpha = 0.6$. $\eta_F$ of the four models decrease with increasing noise intensity, consistent with above regulation, but their $\eta_B$ changes relatively slightly. It is also noteworthy that the SOAE and DSOAE denoisers have external generalization ability, which is illustrated in Figure S7 (Supporting Information).

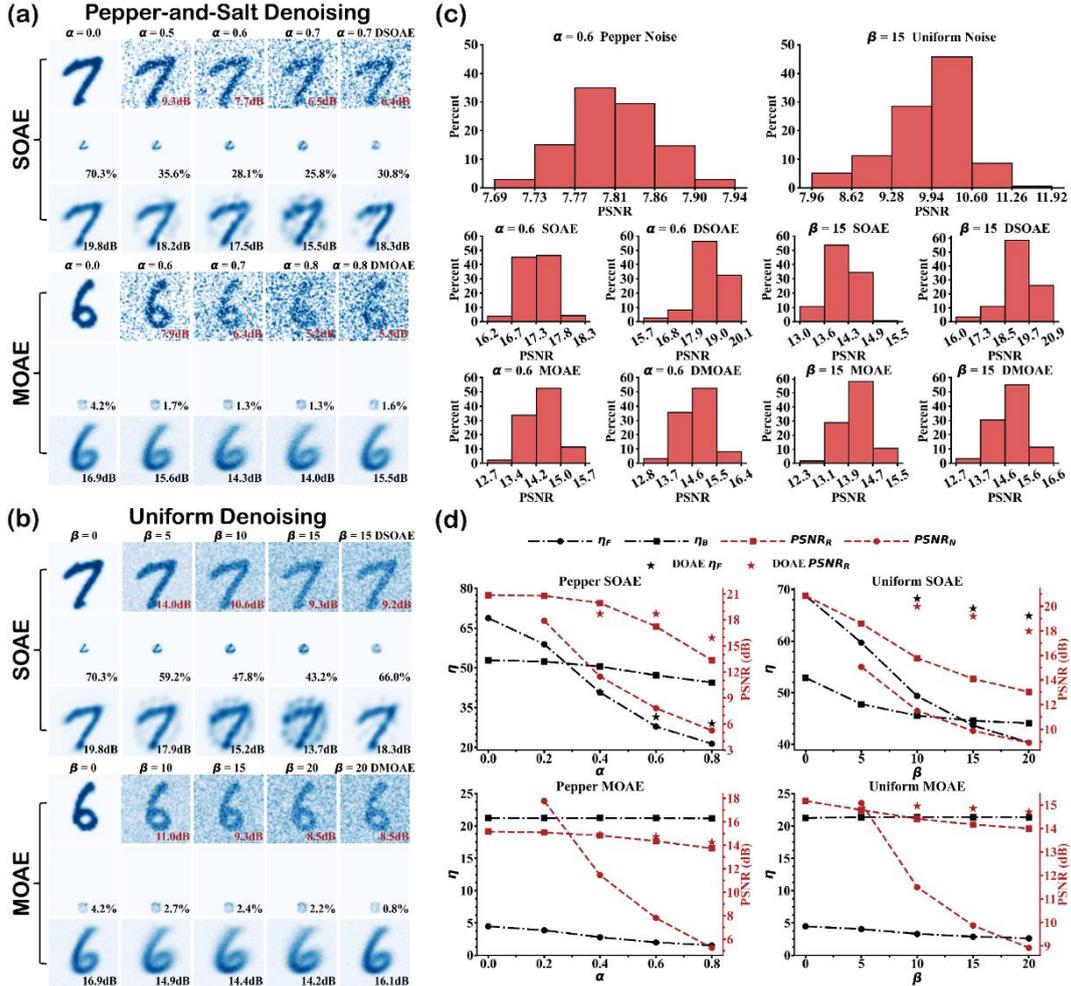

**Figure 3.** The performances of diffractive image denoisers. (a) Results of the SOAE and MOAE models used to remove salt-and-pepper noise from images. The values labeled in images are PSNR and $\eta_F$. DSOAE and DMOAE models are trained by images containing $\alpha = 0.6$ noise. (b) Results of the SOAE and MOAE models used to remove uniformly distributed noise from images. DSOAE and DMOAE models are trained by images containing $\beta = 10$ noise. (c) PSNR distribution histogram of images from the MNIST validation set corrupted by noise with $\alpha = 0.6$ and $\beta = 15$, processed using the SOAE, MOAE, DSOAE, and DMOAE models. (d) Curves showing the variation of transmission efficiency and PSNR with noise intensity when using the SOAE and MOAE models to denoise salt-and-pepper noise and uniformly distributed noise. $PSNR_R$ and $PSNR_N$ are respectively PSNR of reconstructed images and noised images.

**2.3 Application 2: Image classification**

Next, reconfigurable image classifiers are constructed using encoders from well-trained OAE basic models and pluggable phase-only diffractive layers. The encoders with fixed phase profiles function as feature extractors, while the pluggable layers, optimized through training, served for classification. These mechanically reconfigurable classifiers can be trained to handle various datasets by inserting different classification layers after the same encoder. Benefiting from the denoising capability of the encoders, our classifiers also exhibited strong noise resistance, thus we referred to them as noise-resistant reconfigurable image classifiers (NRICs). For convenience, we denote the combination of feature extractors and classification layers using model and dataset names. One example is the SMNIST-1EMNIST NRIC, which represents a classifier trained on EMNIST for classification, including a well-trained SOAE feature extractor from the MNIST dataset and one classification layer.

Combinations of different feature extractors, datasets, and classification layer depths yield 36 NRICs, with their accuracies shown in Figure 4a. Figure S8-S10 (Supporting Information) provide detailed confusion matrices and phase profiles of these NRICs. We analyze the variation in accuracy using 12 NRICs with feature extractors trained on the MNIST dataset as examples. The accuracies of the SMNIST-1MNIST and SMNIST-1EMNIST classifiers are close, being 84.6% and 84.1%, respectively. The SMNIST-1Fashion classifier has the lowest accuracy at 73.2%, due to smaller inter-class differences in the Fashion dataset. The accuracies of SMNIST-2MNIST, SMNIST-2EMNIST, and SMNIST-2Fashion classifiers are improved by approximately 4.5% with the addition of an extra classification layer, reaching 89.7%, 88.1%, and 77.7%, respectively. The classification performances of six NRICs based on the MOAE encoder are similar to the performances of corresponding classifiers mentioned above. The highest accuracies achieved on the MNIST, EMNIST, and Fashion datasets are 89.8%, 88.7%, and 78.4%, respectively, comparable to the accuracies of ordinary diffractive classifiers under the same configurations (see Figure S11**,** Supporting Information). Figure 4b illustrates the impact of noise intensity on the accuracy of the ordinary classifier, SMNIST-2MNIST, and MMNIST-2MNIST NRICs. When images are disturbed by salt-and-pepper noise with $\alpha = 0.6$, the NRIC maintains an accuracy above 81.5%, while that of the ordinary classifier drops to 66.4%. With the increase of the noise intensity, the gaps between the curves get widened, which indicates that the accuracy of ordinary classifier declines more rapidly than the accuracies of NRICs. The NRICs demonstrate even greater resistance to uniformly distributed noise. At $\beta = 0.6$, the accuracies of NRICs are around 79%, while that of ordinary classifier plummets to only 18.9%.

## 2.4 Application 3: Image generation

Finally, the decoders of SOAE and MOAE models are utilized to achieve hologram generations (HGs) and conditional hologram generations (CHGs). For HG, the pre-obtained encoded patterns of original images are directly taken as input patterns for the generators to project holograms. In contrast, for CHG, the generators created holograms utilizing any patterns, with their categories being determined by the positions of the input patterns. Figure 4c illustrates the HGs on MNIST, EMNIST and Fashion datasets. Specifically, we use the average of the encoded patterns for each class of original images as the input patterns. Due to the convex lens-like characteristics of SOAE models trained on MNIST and EMNIST, input patterns of their generators are the average downscaled images of original images, which are then enlarged to generate holograms. Instead, the SOAE generator trained on Fashion dataset and three MOAE generators produce holograms from average encrypted encoded patterns. The generated holograms clearly display the typical features of each category of images but the edges of some images are slightly blurred due to the averaging of multiple encoded patterns.

Figure 4d-4f represent the CHGs using MOAE generator trained on MNIST dataset. By respectively employing ten average encoded patterns at different places of each class as model inputs, image transformations, a specific situation of CHGs was implemented and illustrated in Figure 4d. Each column of Figure 4d presents the result of transforming one digit into other nine digits. The ten categories of images can be flexibly converted into one another by the MOAE generator, and the details of the same category of digits converted from different category digits vary (i.e., The width of 0, the curvature of 2, and the inclinations of 5 and 7). From the perspective of linear space, the ten encoding regions can be considered as the bases of the ten categories of images in the DLS and the pattern content can be thought as the weight of each basis. The forward encoding process decomposes the non-orthogonal original images onto the ten bases in the DLS, while the image transformation process directly utilizes the components on each basis to generate the corresponding category images. Figure 4e and Figure 4f respectively show the CHGs from specific input patterns (constant, gaussian, cosine and sine) and random input patterns (normal distributions and uniform distributions). Even though these input patterns were not from the encoded patterns, images with unique

categorical features were still generated and different patterns produced various image details. It is also noteworthy that the holograms in these two figures contain more noise than holograms in Figure 4d. There are two reasons for this phenomenon. For one thing, our MOAE model can only perform linear operations. While it can separate different categories of images through positional encoding, it struggles to effectively track significant variations within the same category, resulting in only the average output of those images. For another, the content of the encoding patterns for each category lacks a prior distribution, which prevents us from providing the generator with an encoding pattern containing accurate image information. To achieve controllable and clear conditional hologram generation, we can design an optoelectronic hybrid model composed of a nonlinear electronic encoder and a nonlinear $D^2NN^{21}$ generator and train it using the loss function containing positional encoding and prior distribution regularization term. The results of CHGs indicate that a well-trained MOAE model can store the primary information of a dataset in its diffraction layer phase profiles, and input patterns can be utilized as indexes to extract the information. This could provide new insights for enhancing the channel capacity of hypersurface holography and optical storage.

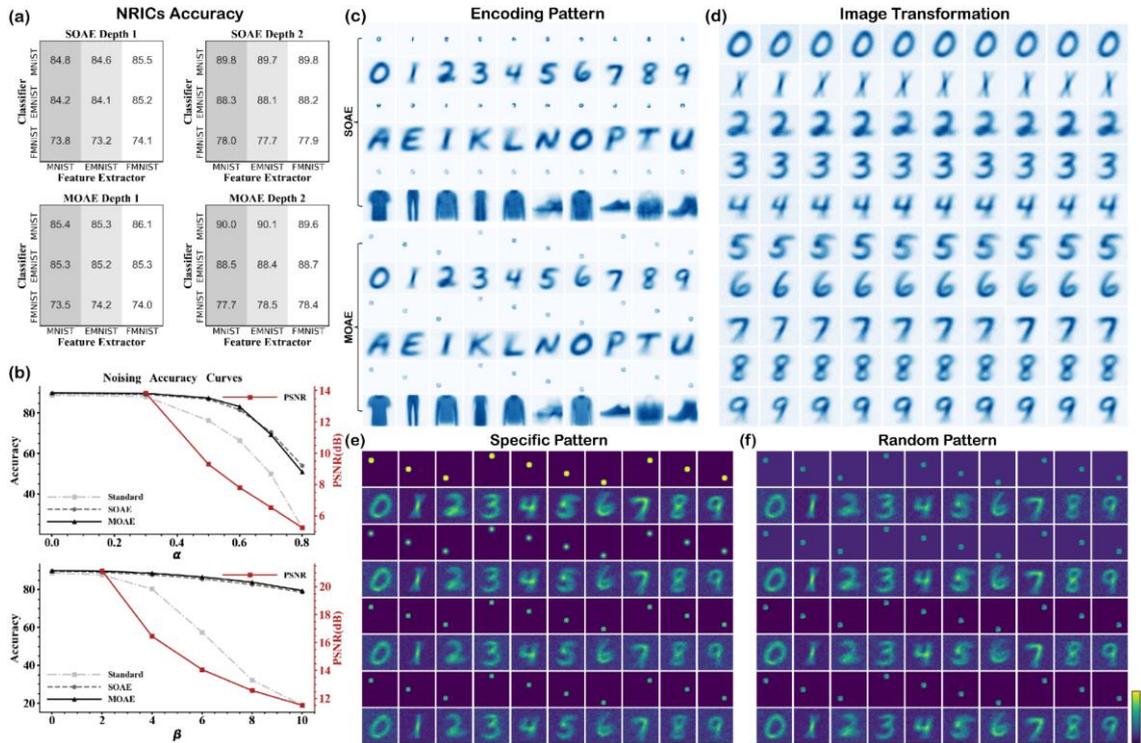

**Figure 4.** The results of NRICs, HGs and CHGs. (a) The accuracy confusion matrices of all 36 types of NRICs (i.e., the combination of Classifier: MNIST and Feature Extractor: EMNIST in SOAE Depth 1 represents the SEMNIST-1MNIST NRIC). (b) The impact curves of noise intensity on classifier accuracy. Standard refers to the normal classifier consisted of seven diffractive layers trained on MNIST dataset, SOAE and MOAE represent

SMNIST-2MNIST and MMNIST-2MNIST NRICs, respectively. (c) Hologram generations using SOAE and MOAE models. 1st, 3rd and 5th rows display average encoded patterns of MNIST, EMNIST and Fashion datasets, respectively. 2th, 4th and 6th rows show the corresponding generated holograms. (d) Image transformations of MNIST dataset based on well-trained MOAE generator (e.g., The transformation of number 0 to number 1 was achieved as follows. Firstly, get ten average encoded patterns of number 0 class. Then, only take the pattern at second place corresponding to number 1 class as input of generator to produce hologram of number 1. See Supplementary Note 3 for details). (e), (f) Conditional hologram generation of MNIST dataset using well-trained MOAE generator.

## 2.5 Validation of the OAE framework

For the purpose of validation, we further trained tiny experimental models on sub-datasets and manufactured them utilizing 3D printing technology. The diffraction layers of the experimental models are composed of periodically arranged micropillars, with the transmission coefficient of each metaatom being controlled by the micropillar height. We designed four experiments to validate the bidirectional transmission properties of the OAE model, including encoding and decoding based on the SOAE model, denoising based on the SOAE model, CHG based on the MOAE model, and NRIC based on the MOAE model. The encoding region of tiny SOAE and MOAE models are both a square with a side length of 8 pixels, resulting in a compression ratio of 16. Note S4 and S7 (Supporting Information) provide the detailed settings of experimental models, sub-datasets and materials.

Figure 5a and 5b respectively illustrate the actual and conceptual experiment systems. A two-stage laser was used to generate 0.69 THz laser light, a Keplerian lens system was employed for beam shaping. The laser was first amplitude modulated by the input sample layer with a hollow pattern (see Figure S14, Supporting Information) to carry the image information, then processed by the OAE model, and finally captured by the THz camera at the imaging plane. Bidirectional transmission in OAE was simulated by swapping the sequence of the difrractive layers (i.e., forward: $L_1$->$L_2$, backward: $L_2$->$L_1$).

Figure 5c presents the experimental verification results of the SOAE model trained on the MNIST subset and its denoising capability. For the convenience of fabrication, we simply binarized the numerically computed encoded patterns (4th column), which were then manufactured for the backward transmission (decoder) experiment. Figure S16 (Supporting Information) further provides two practical designs to directly obtain denoised images through reflection. Due to the quantization loss of the DLS patterns, the reconstructed images have slightly thicker lines than the original images. Figure 5d shows the verification of CHGs using the MOAE generator trained on the MNIST subset. The encoding region distribution of the experimental MOAE model can be found in Figure S17 (Supporting Information). This figure demonstrates that the MOAE model can generate target numbers using constant input patterns at different positions. The height profiles of the tiny 2-layer SOAE and MOAE models are shown in Figure 5e. Figure 5f validates the classification performance of MMNIST-1EMNIST, MMNIST-1MNIST, and MMNIST-1Fashion NRICs. Due to limitations of the experimental platform (i.e., small illumination area of THz camera), we employed an RSC method (see Methods) that utilized fivefold upsampling of the original grids to validate the classification results. The energy distributions calculated by the two methods match very well and the contrasts between maximum energy and second maximum energy are significant, which proves the recognition abilities of NRICs.

Figure S15 (Supporting Information) shows the numerical simulation performance of the experimental models. Figures S15a and S15b indicate that the boundary of the reconstructed images from the binarized DLS using the SOAE model is coarser. The performance metrics in Figure S15c show that the quality of reconstructed images from the two-layer SOAE model and the two-layer MOAE model are lower compared to the five-layer SOAE model and the six-layer MOAE model. Specifically, the SSIM of the former decreases by approximately 0.14, while the SSIM of the latter decreases by approximately 0.06. The SSIM of the binarized SOAE model is 0.06 lower than that of the original SOAE model. Figures S15d-S15e present the denoising results of the two-layer SOAE model and its binarized version. it effectively removes image noise as $\alpha < 0.4$. However, at $\alpha = 0.6$, the denoised images still contain significant noise, and the corresponding binarized model produces images with very low quality. The noise curve shows that the average PSNR of the binarized SOAE model is 1-2 dB lower than that of the original model. Figures S15g-S15h show the CHG results of the MOAE model. It can effectively generate images across five different categories, although the generated digit '2' appears blurrier compared to that from the six-layer MOAE model. Figure S15i displays the confusion matrices for the three MOAE-based NRICs. The accuracies of the three classifiers are 89.8%, 92.0%, and 89.4%, respectively. Besides, we also trained a set of miniature models for CST simulation and their results are presented in Note S8 (Supporting Information).

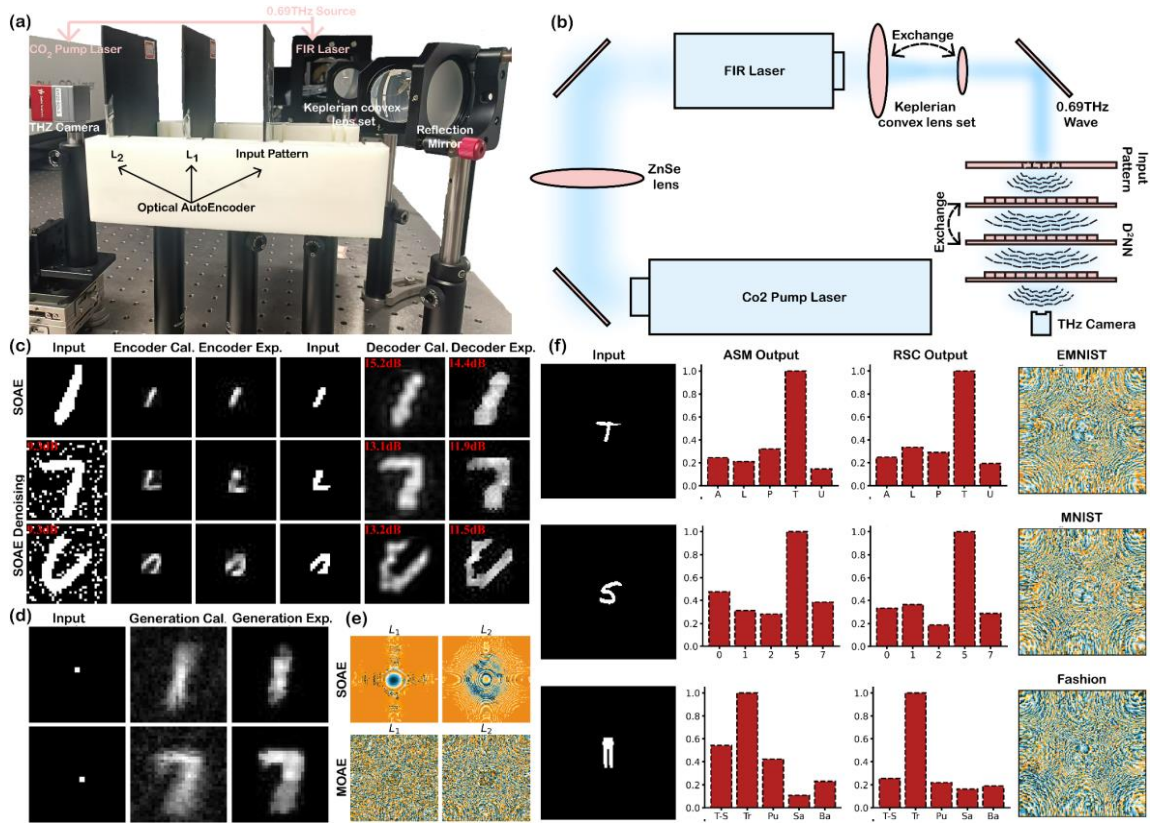

**Figure. 5** The experimental set-up and validation results. (a) Actual optical path diagram of the THz verification experiment. (b) Conceptual schematic diagram of the THz verification experiment. (c) Experimental verification results of the SOAE model. (d) Experimental verification results of the CHGs using MOAE model. (e) Height profiles of the MOAE and SOAE models. (f) Verification results of the NRICs using precise numerical calculation method and their classification layer height profiles.

## 3. Discussion

In this article, we propose an OAE framework for all-optical machine learning. This framework fully exploits the bidirectional information channels of the $D^2NN$, enabling it to function as both an optical encoder and decoder in two transmission directions using only a single set of linear diffractive layers, which enhances the system compactness and integration. By modifying prior shape distributions, various OAE models are developed to demonstrate the encoding and representation capabilities of the DLS. Two basic models, the SOAE and MOAE, both achieve compression encoding with an approximate compression ratio of 52. The SOAE model, utilizing a single encoding region, can learn the imaging principles of convex lenses through the reconstruction of MNIST or EMNIST datasets, which produces high-quality reconstructed images with an SSIM exceeding 0.85. This result showcases the potential of $D^2NNs$ to uncover general physical principles. In contrast, the MOAE model features multiple encoding regions located at different positions, each containing encrypted encoding information for a specific category. It demonstrates the spatial position encoding capability of the DLS. Additionally, six extended modes explore a variety of encoding patterns, further demonstrating that the DLS can flexibly encode raw information into regions of any shape.

The well-trained SOAE and MOAE models are employed to construct image denoisers, classifiers and generators. The dual modulation effect of each diffractive layer, combined with the spatial filtering effect of the compressed encoding in the DLS, enables our denoisers to achieve superior noise reduction performances compared to the unidirectional end-to-end diffractive denoisers[28] with the same number of layers. Additionally, since our denoiser incorporates multi-stage diffraction low-pass filters,[50] it can effectively suppress other types of noise with redundant high-frequency components, such as Gaussian and Poisson noise. By flexibly combining the independently trained encoder (feature extractor) with newly inserted classification layers, our classifiers can achieve classification across multiple datasets in a mechanically reconfigurable manner. Although only three datasets are used in this article, our classifiers can be simply extended to classify more datasets, such as Quickdraw and KMNIST, by inserting new classification layers behind the encoder. Moreover, this assembly approach enables our classifiers to possess stronger noise resistance characteristics compared to conventional diffractive classifiers. Image generators achieve hologram generations and conditional hologram generations. The former directly produces holograms with large FOV from the encoded patterns in small regions, which can be used in AR or VR.[42] The latter projects different categories of images based on positions of input patterns. The conditional hologram generator indicates that a well-trained MOAE model can be regarded as a special memory that retains dataset information, with its index defined by the position and content of the input patterns. This will offer a new approach for enhancing the capacity of metasurface holography and optical information storage.[43-45] Furthermore, the encrypted characteristics and shape-controllable features of the encoding patterns in the six extended modes make them promising candidates for optical image encryption and obstacle-avoiding optical communication.[38,46,47]

The OAE framework also exhibits certain limitations. Due to the energy loss of transmissive metasurfaces, the reconstruction efficiency of systems requiring bidirectional transmission mechanisms (e.g., Image denoising) will significantly decrease with the additional transmissive modulation. Carefully designed high-transmittance metasurfaces[48,49] can help improve system efficiency to a certain extent. Besides, the input and output FOV of the bidirectional transmission system are the same, which hinders the measurement of output results. We illustrate two practical designs using beam splitters or off-axis imaging in Figure S16 (Supporting Information) to solve the problem. Finally, since our OAE models only contain linear phase modulation and linear diffraction processes, the DLS inherently implements latent representations of data through linear transformations. The linear nature makes them only applicable to simple-featured MNIST-like datasets, while exhibiting significant performance degradation on complex datasets containing richer detailed information such as CIFAR-10 and ImageNet. The introduction of nonlinear operations into the OAE model will mitigate this limitation.

In summary, our OAE framework explores new dimensions of optical encoding in shape encoding and positional encoding, characterized by compactness and multifunctional integration. It can be applied in various fields such as pattern recognition, image processing, computational holography, optical storage, optical communication, and optical image encryption, and is expected to promote the development of general intelligent optical systems.

## 4. Methods

*Numerical propagation model for diffractive processors:* The diffractive processor comprised an input layer, a set of diffractive layers and an output layer, each of which respectively located at a different axial position $z_l$ ($l = 0, 1, 2, ... N, N + 1$). The input wavefield can be modulated layer by layer when propagating from input plane ($z_0$) to output plane ($z_{N+1}$). Modulation effect of the $l$ th layer ($1 \leq l \leq N$) was expressed as follows:

$$U_l^o(x, y) = U_l^i(x, y) T_l(x, y)$$
$$T_l(x, y) = a_l(x, y) \exp(j\phi_l(x, y)) \quad (1)$$

where $U_l^i(x, y)$ and $U_l^o(x, y)$ respectively represent input and output wavefield of the $l$ th layer. $T_l(x, y)$ is transmission coefficient of $l$ th layer.

Propagation process between the $l$ th layer and the $l + 1$ th layer ($0 \leq l \leq N$) can be described by the Rayleigh Sommerfeld convolution formula (RSC)[50] as follows:

$$U_{l+1}^i(x, y) = U_l^o(x, y) * W(x, y, d_l)$$
$$d_l = z_{l+1} - z_l$$
$$W(x, y, d_l) = \frac{d_l}{r^2} \left(\frac{1}{2\pi r} + \frac{1}{j\lambda}\right) exp\left(j\frac{2\pi r}{\lambda}\right)$$
$$r = \sqrt{x^2 + y^2 + d_l^2} \quad (2)$$

where $d_l$ is the distance between two layers, and $W(x, y, d_l)$ is the convolution operator that reflects the diffraction connection of two layers.

Even though the RSC method is accurate enough to describe the diffraction process between two layers, the overlarge computational effort caused by large-kernel convolution made it impossible to realize in the training procedure. Therefore, we only used this method for numerical validation. Instead, a lower-computing and faster algorithm called angular spectrum method (ASM) was utilized for model training and inference, which can be expressed as follows:

$$U_{l+1}^i(x,y) = IFFT(FFT(U_l^o(x,y))H(f_x, f_y, d_l))$$

$$H(f_x, f_y, d_l) = \exp\left(\frac{j2\pi d_l}{\lambda}\sqrt{1-(\lambda f_x)^2-(\lambda f_y)^2}\right) \quad (3)$$

where $H(f_x, f_y, d_l)$ is the spectrum transfer function, $f_x$ and $f_y$ are spatial frequencies along x and y axis, respectively.

*Loss functions and metrics:* The OAE models function as encoders when input images pass through them in the forward direction ($FOV_\mathrm{I} \rightarrow FOV_\mathrm{II}$), Conversely, they act as decoders when the encoded patterns pass through in the backward direction ($FOV_\mathrm{II} \rightarrow FOV_\mathrm{I}$). We wrote the Loss function for training OAE models as follows:

$$\mathcal{L}_{OAE} = \mathcal{L}_{Rec} + \gamma_{En}\mathcal{L}_{En} + \gamma_{Eff}\mathcal{L}_{Eff} \quad (4)$$

Here, $\mathcal{L}_{Rec}$ represents the reconstruction error between input images and corresponding decoded images, and is defined as:

$$\mathcal{L}_{Rec} = \mathcal{L}_{MSE}(I(x,y), O_{De}(x,y)) = E[|\sigma_1 I(x,y) - \sigma_2 O_{De}(x,y)|^2]$$

$$\sigma_1 = \frac{1}{\sum_{(x,y)\in FOV_\mathrm{I}} I(x,y)}$$

$$\sigma_2 = \sigma_1 \frac{\sum_{(x,y)\in FOV_\mathrm{I}} I(x,y)O_{De}(x,y)}{\sum_{(x,y)\in FOV_\mathrm{I}} |O_{De}(x,y)|^2} \quad (5)$$

where $E[\cdot]$ is the average operator, $I(x,y)$ stands the input image and $O_{De}(x,y)$ stands the decoded image. $\sigma_1$ and $\sigma_2$ are parameters that eliminate the error due to diffractive energy loss.

$\mathcal{L}_{En}$ is a regularization term designed to maximize the concentration of energy in the target encoding region, which can be expressed as:

$$\mathcal{L}_{En} = -\ln\left(\frac{\sum_{(x,y)\in FOV_\mathrm{II}} |M(x,y)O_{En}(x,y)|^2}{\sum_{(x,y)\in FOV_\mathrm{II}} |O_{En}(x,y)|^2}\right)$$

$$M(x,y) = \begin{cases} 1, & (x,y) \in encoding\ region \\ 0, & otherwise \end{cases} \quad (6)$$

where $O_{En}(x, y)$ represents the encoder output in $FOV_{II}$. $M(x, y)$, a region mask function that defines the shape of encoding patterns, guides the output to align with the target shape by providing a prior shape distribution, similar to the prior probability distribution in the electronic VAE model.

$\mathcal{L}_{Eff}$ is another regulation term that controls the bidirectional diffraction efficiency. For SOAE and its extended models, we simply used the reconstruction efficiency to constrain the model weights. The equation can be expressed as follows:

$$\mathcal{L}_{Eff} = \begin{cases} -\ln\left(\frac{\eta_R}{\eta_{Th}}\right), & \eta_R < \eta_{Th} \\ 0, & \eta_R \geq \eta_{Th} \end{cases}$$

$$\eta_R = \frac{\sum_{(x,y)\in FOV_I} |O_{De}(x,y)|^2}{\sum_{(x,y)\in FOV_I} |I(x,y)|^2} \tag{7}$$

In the MOAE model, both forward and backward efficiency were used to more accurately regulate the bidirectional efficiency, as expressed below:

$$\mathcal{L}_{Eff} = \begin{cases} -\ln\left(\frac{\eta_F}{\eta_{Fth}}\right), & \eta_F < \eta_{Fth} \\ 0, & \eta_F \geq \eta_{Fth} \end{cases} + \begin{cases} -\ln\left(\frac{\eta_B}{\eta_{Bth}}\right), & \eta_B < \eta_{Bth} \\ 0, & \eta_B \geq \eta_{Bth} \end{cases}$$

$$\eta_F = \frac{\sum_{(x,y)\in FOV_{II}} |M(x,y)O_{En}(x,y)|^2}{\sum_{(x,y)\in FOV_I} |I(x,y)|^2}, \quad \eta_B = \frac{\sum_{(x,y)\in FOV_I} |O_{De}(x,y)|^2}{\sum_{(x,y)\in FOV_{II}} |M(x,y)O_{En}(x,y)|^2} \tag{8}$$

$\eta_{Th}$, $\eta_{Fth}$ and $\eta_{Bth}$ are hyperparameters that control the total, forward, and backward efficiencies, respectively. Similarly, $\gamma_{En}$ and $\gamma_{Eff}$ in equation (4) are hyperparameters representing the weight coefficients for the encoding shape and efficiency penalty terms.

Moreover, The encoders ($FOV_I \rightarrow FOV_{II}$) from trained SOAE and MOAE models were further utilized for classification. They were taken as generalized feature extractors, capturing compact information from input images. By cascading one or two to-be-trained diffractive layers behind the fixed encoders, we implement $\mathcal{L}_{\text{NRIC}}$ using $\mathcal{L}_{MSE}$ in equation (5) to train image classifiers. The formula of $\mathcal{L}_{\text{NRIC}}$ was as follows:

$$\mathcal{L}_{\text{NRIC}} = \mathcal{L}_{MSE}\big(S(x,y),\, O_{\text{NRIC}}(x,y)\big) \tag{9}$$

where $S(x, y)$ represents the focus pattern associated with the input class, $O_{\text{NRIC}}(x, y)$ denotes the output wavefield of classifier.

In addition to the above-mentioned MSE and efficiency ($\eta$), we also used two additional metrics: SSIM and PSNR, to comprehensively evaluate the performance of the trained models. SSIM measured the similarity between images based on brightness, contrast, and structure, and can be expressed as follows:

$$SSIM = \frac{(2E[I(x,y)]+C_1)(2Cov[I(x,y),O_{De}(x,y)]+C_2)}{(E[I(x,y)]^2+E[O_{De}(x,y)]^2+C_1)(Var[I(x,y)]+Var[O_{De}(x,y)]+C_2)} \tag{10}$$

Here, $Cov[\cdot,\cdot]$ calculates the covariance between two images, while $Var[\cdot]$ represents the variance of an image. $C_1$ and $C_2$ are respectively 0.01 and 0.03. To more accurately assess similarity, we divided the two images into multiple blocks of size W×W and compute the average SSIM across these blocks as the final results.

PSNR quantified the noise in the reconstructed image compared to the original image and can be expressed as follows:

$$PSNR = 10 log_{10} \frac{1}{E\left[\left|\tilde{I}(x,y)-\widetilde{O_{De}}(x,y)\right|^2\right]} \tag{11}$$

where ~ is normalization operator.

**Supporting Information**

Supporting Information is available from the Wiley Online Library or from the author.

**Acknowledgements**

The authors acknowledge the financial support by the National Natural Science Foundation of China (Grants No.61991423, 62205294, and 52332006), the National Key R&D Program of China (2022YFB3806000), and Beijing Natural Science Foundation (Z240008).